\documentclass{aa}
\usepackage{graphicx}
\usepackage{txfonts}
\newcommand{\etal}{et\,al.\ }
\newcommand{\logg}{\mbox{$\log g$}}
\newcommand{\Teff}{\mbox{$T_\mathrm{eff}$}}
\newcommand{\hh}{\object{H\,1504$+$65}}
\newcommand{\alp}{\object{$\alpha$\,Cen}}
\newcommand{\ala}{\object{$\alpha$\,Cen\,A}}
\newcommand{\alb}{\object{$\alpha$\,Cen\,B}}
\newcommand{\alab}{\object{$\alpha$\,Cen\,A and B}}
\newcommand{\pp}{\object{Procyon}}
\newcommand{\hse}{\hbox{}\hspace{1.1mm}\hbox{}}
\newcommand{\hsw}{\hbox{}\hspace{1.3mm}\hbox{}}
\newcommand{\hsed}{\hbox{}\hspace{0.8mm}\hbox{}}
\newcommand{\hses}{\hbox{}\hspace{0.4mm}\hbox{}}
\newcommand{\hsx}{\hbox{}\hspace{1.5mm}\hbox{}}
\begin{document}
   \title
   {
Line identification in soft X-ray spectra of stellar coronae by comparison with
   the hottest white dwarf's photosphere:\\ Procyon, $\alpha$\,Cen~A+B, and \hh}
 
   \author{K. Werner$^1$ and J.J. Drake$^2$}
   \offprints{K\@. Werner}
   \mail{werner@astro.uni-tuebingen.de}
 
   \institute
    {
     Institut f\"ur Astronomie und Astrophysik, Universit\"at T\"ubingen, Sand 1, 72076 T\"ubingen, Germany
\and
Harvard-Smithsonian Center for Astrophysics, MS 3, 60 Garden Street, Cambridge,
    MA 02138, USA
}
    \date{Received xxx / Accepted xxx}
   \authorrunning{K. Werner \& J.J. Drake}
   \titlerunning{Line identification in soft X-ray spectra of stellar coronae}
   \abstract{\hh\ is a young white dwarf with an effective temperature of
200\,000\,K and is the hottest post-AGB star ever analysed with detailed model
atmospheres. {\it Chandra} LETG+HRC-S spectra have revealed the richest X-ray
absorption line spectrum recorded from a stellar photosphere to date. The line
forming regions in this extremely hot photosphere produce many transitions in
absorption that are also observed in emission in cool star coronae. We have
performed a detailed comparison of {\it Chandra} spectra of \hh\ with those
of Procyon and $\alpha$\,Cen~A and B. State of the art non-LTE model spectra for
the hot white dwarf have enabled us to identify a wealth of absorption lines
from highly ionized O, Ne and Mg. In turn, these features have allowed us to
identify coronal lines whose origins were hitherto unknown.
        }
   \maketitle

\section{Introduction} 

High-resolution X-ray spectroscopy performed with {\it Chandra} and XMM-Newton
allows very detailed studies of coronae about cool stars. While many individual
emission lines were detected for the first time in stellar spectra by the
Extreme Ultraviolet Explorer Satellite (EUVE; see, e.g., Drake \etal 1995), the
resolving power of $\lambda/\Delta\lambda\sim 200$ of the EUVE spectrographs was
a quite modest compared with that of present day X-ray observatories.  In
particular, the unprecedented resolution capabilities of the {\it Chandra} X-ray
Observatory Low Energy Transmission Grating Spectrograph (LETG) in the
30-170~\AA\ range ($\lambda/\Delta\lambda\sim 1000$) that overlaps with the EUVE
Short Wavelength spectrometer (70-170~\AA), have revealed many more weak
spectral lines.

The 25-70~\AA\ region is a relatively uncharted part of the soft X-ray spectrum.
Prior to {\it Chandra}, only a small handful of astrophysical observations had
been made at anything approaching high spectral resolution in this range: these
were of the solar corona using photographic spectrometers (Widing \& Sandlin
1968; Freeman \& Jones 1970; Schweizer \& Schmidtke 1971; Behring \etal 1972; 
Acton \etal 1985) a channel electron photomultiplier (Malinovsky
\& Heroux 1973) and a Geiger-M\"uller counter (Manson 1972).  While these works
resulted in identifications for many of the bright spectral lines, a large
fraction of the forest of weaker features remains unidentified.  Identification
of these features is desirable because they could be used as spectroscopic
diagnostics, because they potentially contribute to the flux of diagnostic lines
currently employed, and because they contribute to the overall plasma radiative
loss.

Two nearby stars that have illuminated the forest of lines in the 30-170~\AA\
range are \alp\ (G2V+K1V) and \pp\ (F5IV).  All three stars exhibit classical
solar-like X-ray emitting coronae.  Indeed, analogues of the relatively X-ray
faint Sun are difficult to observe because they become unreachable with current
instrumentation beyond a few parsecs, and \alp\ and \pp\ represent the nearest
and brightest coronal sources with solar-like activity.  Only a small fraction
of the multitude of lines between 30-170~\AA\ seen in their {\it Chandra} LETG
spectra could be identified based on current radiative loss models (Raassen
\etal\ 2002, 2003).  Drake \etal (in prep.) have estimated that these models
underestimate the true line flux in the range 30-70~\AA\ in these stars by
factors of up to 5 or so.

The ``missing lines'' are predominantly transitions involving $n=2$ ground
states in abundant elements such as Ne, Mg, Si, S and Ar---the analogous
transitions to the Fe ``L-shell'' lines between $\sim 8$-18~\AA, together with
Fe $n=3$ (the ``M-shell'') transitions (Drake 1996, Drake et al.\ 1997, Jordan
1996).  Some of these lines have been identified based on Electron Beam Ion Trap
experiments (Beiersdorfer \etal 1999, Lepson \etal 2002, 2003).  In the present
paper we approach this problem from a new perspective, namely through a {\it
Chandra} observation of the photosphere of the hottest white dwarf (WD) known,
\hh, and its quantitative analysis by means of detailed non-LTE model
atmospheres.

\hh\ has an effective temperature of 200\,000\,K. It belongs to the PG1159
spectral class, which are hot, hydrogen-deficient (pre-) white dwarfs. Their
surface chemistry (typical abundances: He=33\%, C=48\%, O=17\%, Ne=2\%, mass
fractions) suggests that they exhibit matter from the helium-buffer layer
between the H- and He-burning shells in the progenitor AGB star (Werner
2001). This is likely because the PG1159 stars have suffered a late He-shell
flash, a phenomenon that drives the fast evolutionary rates of such famous stars
like FG\,Sge and Sakurai's object.  \hh\ is in fact a peculiar member of this
class, because it is also helium-deficient.  Its atmosphere is mainly composed
of carbon and oxygen plus neon and magnesium (C=48\%, O=48\%, Ne=2\%, Mg=2\%,
mass fractions).  \hh\ is a unique object, considering its high \Teff\ and
chemical surface composition, and we have speculated that it represents the
naked C/O core of a former red giant (Werner \etal 2004, W04).

{\it Chandra} LETG+HRC-S spectra from \hh\ have revealed the richest X-ray
absorption line spectrum recorded from a stellar photosphere to date. We have
recently performed a detailed analysis of this spectrum (W04) and we use in the
paper in hand the photospheric spectrum of \hh\ together with an appropriate
model atmosphere to identify a number of emission lines in the coronae of \ala,
\alb, and Procyon. The difference in particle densities in the WD photosphere
and in the coronae amounts to many orders of magnitude  (roughly n$_{\rm
e}$=$10^{13}-10^{18}$ and $10^{10}$\,cm$^{-3}$, respectively),  however, the
temperature in the line forming regions of the WD (up to 300\,000\,K) is
comparable to the low-temperature component of multi-temperature fits to
coronae, required to account for the lines of low-ionization stages (e.g.\
630\,000\,K for Procyon; Raassen \etal 2002). As a consequence, numerous lines
from \ion{O}{vi}, \ion{Ne}{vi-viii} and \ion{Mg}{vi-ix} are visible in the soft
X-ray spectra of both, the cool star coronae (in emission) and the hot WD
photosphere (in absorption). Lines from higher ionization stages are formed in
the high-temperature regions of the coronae (T of the order 1--2.5 million K for
the stars studied in this paper), hence, their respective absorption line
counterparts cannot be formed in the WD photosphere.

In the following, we first introduce briefly the characteristics of the objects
studied here.  We describe our model atmosphere calculation for the hot WD,
concentrating on the atomic data employed.  We then perform a detailed
comparison of the absorption and emission line spectra and suggest a number of
new line identifications for the cool star coronae.

\begin{figure}[tbp]
  \resizebox{\hsize}{!}{\includegraphics{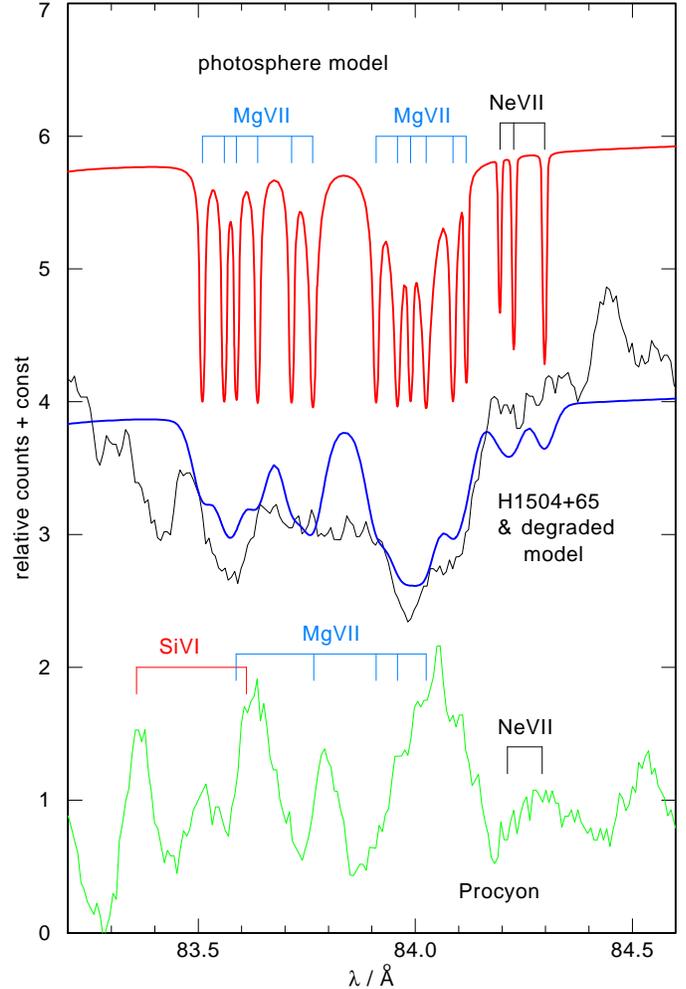}} \caption[]{ Comparison of
  {\it Chandra} X-ray spectra of \hh\ and Procyon. Lines from \ion{Mg}{vii} and
  \ion{Ne}{vii} are in absorption in \hh\ and in emission in Procyon. Top:
  photosphere model for \hh\ with line identifications for \ion{Mg}{vii} and
  \ion{Ne}{vii}. Middle: Degraded model spectrum (i.e.\ folded with a 0.05~\AA\
  FWHM Gaussian) plotted over \hh\ observation. Bottom: Procyon spectrum with
  line identifications from Raassen \etal (2002). {\it Chandra} spectra were
  smoothed with a 0.03~\AA\ boxcar.  } \label{fig_84}
\end{figure}

\begin{figure*}[tbp]
\includegraphics[width=\textwidth]{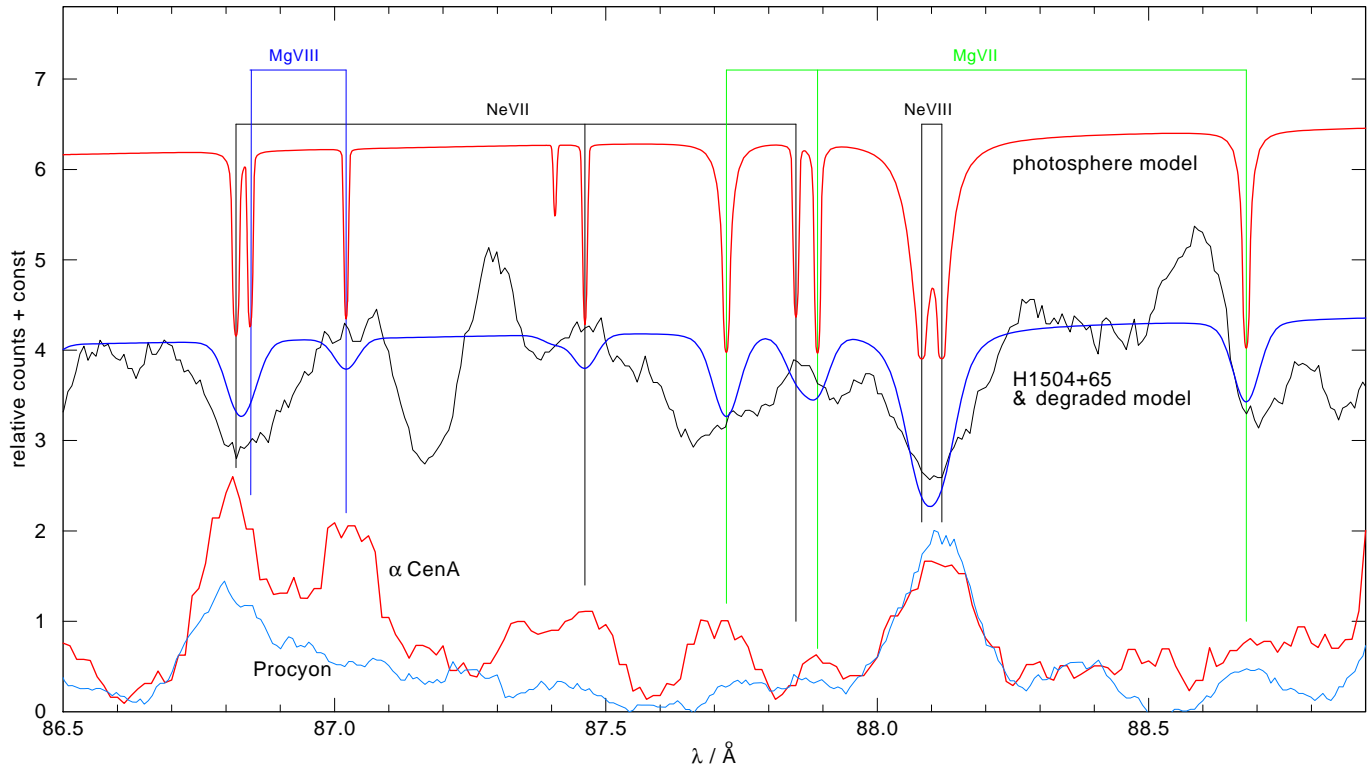} \caption[]{Comparison of {\it
  Chandra} X-ray spectra of \hh\ with Procyon and \ala, similar to Fig.\,1. All
  shown lines from highly ionized Ne and Mg are identified for the first time in
  the cool star corona, except for \ion{Mg}{viii}~86.85/87.02~\AA\ and the
  \ion{Ne}{viii} doublet at 88.1~\AA, which were identified by Raassen \etal
  (2002, 2003). {\it Chandra} spectra of \hh\ and the coronae were smoothed with
  0.03~\AA\ and 0.05~\AA\ boxcars, respectively.  } \label{fig_88}
\end{figure*}

\begin{table*}
\caption{List of X-ray multiplets in the wavelength region 69--151\,\AA\ observed in both the \hh\
photosphere or its model and in the coronae of either \ala\ (``A''), \alb\
(``B''), or \pp\ (``P''), as suggested in this paper. In the last column, we
note earlier line identifications in either solar spectra (SA=Acton et al.\
1985; SB=Behring et al.\ 1972; SF=Freeman \& Jones 1970; SM=Manson 1972;
SMH=Malinovsky \& Heroux 1973; SW=Widing \& Sandlin 1968) or stellar spectra
(D=Drake et al.\ 1995; R=Raassen \etal 2002, 2003).  The letter ``u'' is
appended in the case of the feature having been observed but not
identified. ``N'' denotes a new identification suggested in this paper. ``N'' in
combination with other letters means that at least one component of the
multiplet is newly identified here. Expressions in brackets denote doubtful
cases. The column ``Source'' gives the reference to the level energies of the
transition. After each transition we have marked, if the lower level is a ground
state (``G'') or a metastable state (``M'').
\label{lines_tab}
}
\begin{tabular}{lllr@{\,\,--\,\,}llll}
      \hline
      \hline
      \noalign{\smallskip}
$\lambda$/\AA\ (\hh\ model)     & Seen in  & Ion & \multicolumn{2}{c}{Transition} & & Source & Remark \\
      \noalign{\smallskip}
     \hline
      \noalign{\smallskip}
69.41, .47, .57                & A,B,P  & \ion{Mg}{viii} &
      2p\hse\,$^2$P$^{\rm o}$  & 3p\hse\,$^2$D\hse       &G & {\sc
      Nist} & N,SMu, blend with blue wing of\\
  & & & & & & & \ion{Si}{viii}~69.63~\AA\\
      
74.27, .32, .34, .37, .41, .43 & A,B,P  & \ion{Mg}{viii} & 2p$^2$\,$^4$P\hse        & 3d\hse\,$^4$D$^{\rm o}$ &M & {\sc Nist} & N,SAu,SMu, broad emission feature\\
74.78, .81, .87                & A,B    & \ion{Ne}{vii}  & 2p\hsed\,$^3$P$^{\rm o}$ & 4p\,$^3$D\hse           &M & {\sc Nist} & N,SAu,SBu,SMu, blend with \ion{Mg}{viii}~74.86~\AA\\
74.86, 75.03, .04              & A,B,P  & \ion{Mg}{viii} &
2p\hse\,$^2$P$^{\rm o}$  & 3d\hse\,$^2$D\hse       &G & {\sc Nist} &
SA,SB,SF,SM,SMH,SW,R, blend with \\
  & & & & & & & \ion{Fe}{xiii}~74.85 noted by R, and \ion{Ne}{vii}~74.87~\AA\\
(78.34), 78.41, 78.52          & A,B,P  & \ion{Mg}{vii}  & 2p$^3$\,$^3$P\hse        & 3p\hse\,$^3$P$^{\rm o}$ &G & Kelly      & N,SAu,SMu\\
80.23, .25                     & A,B,P  & \ion{Mg}{viii} & 2p$^2$\,$^2$D\hses       & 3d\hse\,$^2$D$^{\rm o}$ &  & {\sc Nist} & SMu,R\\
80.95, 81.02, .14              & A,(B,P)& \ion{Mg}{vii}  & 2p$^3$\,$^3$P\hse        & 3p\hse\,$^3$S$^{\rm o}$ &G & Kelly      & N,SAu,SMu\\
81.37                          & (A),B,P& \ion{Ne}{vii}  & 2p\hsw\,$^1$P$^{\rm o}$  & 4p\,$^1$P\hse           &  & Kelly      & N\\
81.73, .79, .84, .87, .94, .98 & A,B,P  & \ion{Mg}{viii} &
2p$^2$\,$^4$P\hse        & 3s\hse\,$^4$P$^{\rm o}$ &M & {\sc Nist} &
N,SFu,SMu, broad emission, blend with \\
  & & & & & & & \ion{Si}{vii}~81.89~\AA\ noted by R\\
82.17, .20, .27                & A,B,(P)& \ion{Ne}{vii}  & 2p\hse\,$^3$P$^{\rm o}$  & 4d\,$^3$D\hse           &M & Kelly      & N\\
(82.60), .82                   & A,B,P  & \ion{Mg}{viii} & 2p\hse\,$^2$P$^{\rm o}$  & 3s\hse\,$^2$S\hse       &G & {\sc Nist} & (SBu,SFu),SM,R, blend with \ion{Fe}{xii} 82.84~\AA\ \\
  & & & & & & & noted by R\\
83.51, .56, .59, .64, .71, .76 & A,B,P  & \ion{Mg}{vii}  &
2p$^3$\,$^3$P\hse        & 3d\hse\,$^3$P$^{\rm o}$ &G & Kelly      &
N, SBu,SF,SW,SM,D,R, broad emission\\
  & & & & & & & feature; poss.~\ion{Si}{VI} contribution noted by R\\
83.91, .96, .99, 84.02, .09, .11&A,B,P  & \ion{Mg}{vii}  &
2p$^3$\,$^3$P\hse        & 3d\hse\,$^3$D$^{\rm o}$ &G & Kelly      &
N, SB,SF,SM,SMH,R, broad\\
  & & & & & & & emission feature\\
(84.19, .23,) .30              & A      & \ion{Ne}{vii}  & 2p\hse\,$^3$P$^{\rm o}$  & 4s\,$^3$S\hse           &M & Bashkin    & SMu,R\\
85.41                          & (A,B,P)& \ion{Mg}{vii}  &
2p$^2$\,$^1$D\hses       & 3d\hse\,$^1$F$^{\rm o}$ &M & Kelly      &
N, SBu,SMu, blend with \ion{Fe}{xii}~85.46~\AA\ \\
  & & & & & & & noted by R\\
86.82                          & A,B,P  & \ion{Ne}{vii}  &
2p$^2$\,$^1$D\hses       & 4d\,$^1$F$^{\rm o}$     &M & Kelly      &
N, SBu,SFu,SMu, blend with \ion{Fe}{xi}~86.77, \\
  & & & & & & & \ion{Mg}{viii}~86.84~\AA\ noted by R\\ 
86.84, .85, 87.02              & A,B,P  & \ion{Mg}{viii} &
2p$^2$\,$^2$D\hses       & 3s\hse\,$^2$P$^{\rm o}$ &  & {\sc Nist} &
N, SBu,SFu,SMu,R\\
87.46                          & A      & \ion{Ne}{vii}  & 2s$^2$\,$^1$S\hsed       & 3s\,$^1$P$^{\rm o}$     &G & {\sc Nist} & N\\
87.72                          & A      & \ion{Mg}{vii}  & 2p$^2$\,$^1$D\hses       & 3d\hse\,$^1$D$^{\rm o}$ &M & Kelly      & N\\
88.08, 88.12                   & A,B,P  & \ion{Ne}{viii} & 2s\hse\,$^2$S\hse        & 3p\,$^2$P$^{\rm o}$     &M & {\sc Nist} & SA,SB,SF,SM,SMH,D,R\\
88.68                          & (A),B,P& \ion{Mg}{vii}  &
2p$^2$\,$^1$S\hsed       & 3d\hse\,$^1$P$^{\rm o}$ &M & Kelly      &
N, SMu\\
89.64, .65                     & A,(P)  & \ion{Mg}{vi}   &
2p$^3$\,$^2$P$^{\rm o}$  & 4s\hse\,$^2$P           &M & Kelly      &
N, SBu\\
91.56                          & P      & \ion{Ne}{vii}  & 2p\hsw\,$^1$P$^{\rm o}$  & 4s\,$^1$S\hse           &  & Kelly      & SMu,R\\
92.13, .32                     & A,B,P  & \ion{Mg}{viii} &
2p$^2$\,$^2$S\hsed       & 3s\hse\,$^2$P$^{\rm o}$ &  & {\sc Nist} &
N, SMu,SBu,R\\
92.85                          & P      & \ion{Ne}{vii}  & 2p$^2$\,$^1$S\hsed       & 4d\,$^1$P$^{\rm o}$     &M & Kelly      & SMu,R\\
(93.89), 94.07, .10, (.27)     & A,B,P  & \ion{Mg}{viii} &
2p\hse\,$^2$P\hse        & 3s\hse\,$^2$P$^{\rm o}$ &  & {\sc Nist} &
N, (SAu,SBu),SMu, blend with  \\
  & & & & & & & \ion{Fe}{x}~94.012~\AA, \ion{Mg}{vii}~94.04~\AA\\
94.04, (.17, .24)              & A,B,P  & \ion{Mg}{vii}  &
2p$^3$\,$^5$S$^{\rm o}$  & 3s\hse\,$^5$P\hse       &M & Kelly      &
N, SMu,SFu, blend with \ion{Fe}{x}~94.012~\AA \\
  & & & & & & & noted by R, and \ion{Mg}{viii}~94.07~\AA\\
           \noalign{\smallskip}
\hline
     \end{tabular}
\end{table*}

\addtocounter{table}{-1}
\begin{table*}
\caption{continued}
\begin{tabular}{lllr@{\,\,--\,\,}llll}
      \hline
      \hline
      \noalign{\smallskip}
$\lambda$/\AA\ (\hh\ model)     & Seen in  & Ion & \multicolumn{2}{c}{Transition} & & Source & Remark \\
      \noalign{\smallskip}
     \hline
      \noalign{\smallskip}
94.26, .27, .30, .31, .36, .39 & B      & \ion{Ne}{vii}  &
2p\hse\,$^3$P$^{\rm o}$  & 3p\,$^3$P\hse           &M & Bashkin    &
N, SMu\\
95.03, .04                     & B      & \ion{Mg}{vii}  & 2p$^3$\,$^3$D$^{\rm o}$  & 3s'\,$^3$D              &  & Kelly      & N\\
95.26, .38, .42, .49, .56, .65 & (A,B,P)& \ion{Mg}{vii}  &
2p$^3$\,$^3$P\hse        & 3s\hse\,$^3$P$^{\rm o}$ &G & Kelly      &
N, SBu,SFu,SMu, blend with \ion{Fe}{x}~95.338~\AA\ \\
  & & & & & & & noted by R; \ion{Mg}{vi}~95.42~\AA\\
(95.38, .42, .48)              & (A,B,P)& \ion{Mg}{vi}   & 2p$^3$\,$^4$S$^{\rm o}$  & 3d\hse\,$^4$P           &G & Kelly      & SBu,SFu,SMu,R, blend with \\
  & & & & & & & \ion{Mg}{vii}~95.26--.65~\AA\\
95.75, .81, .89, .90, .91, 96.0&A,B,P   & \ion{Ne}{vii}  &
2p\hse\,$^3$P$^{\rm o}$  & 3p\,$^3$D\hse           &M & Bashkin    &
N, SMu, broad emission, blend with \\
  & & & & & & & \ion{Si}{vi}~96.02~\AA\ noted by R\\
96.08, .09                     & (A,B,P)& \ion{Mg}{vi}   &
      2p$^3$\,$^2$P$^{\rm o}$  & 3d''\,$^2$D             &M & Kelly
      & N, SBu,SMu, blend with \ion{Fe}{x}~96.12~\AA\\
  & & & & & & & noted by R\\
97.50                          & A,B,P  & \ion{Ne}{vii}  & 2s$^2$\,$^1$S\hse        & 3p\,$^1$P$^{\rm o}$     &G & Kelly      & SM,SW,SMH,R\\
98.11, .26                     & A,B,P  & \ion{Ne}{viii} & 2p\hse\,$^2$P$^{\rm o}$  & 3d\,$^2$D\hse           &  & {\sc Nist} & SB,SM,SMH,SW,D,R\\
98.50, .51                     & B      & \ion{Mg}{vi}   &
      2p$^3$\,$^2$P$^{\rm o}$  & 3d'\,$^2$S              &M & Kelly
      & N, SBu,SMu\\
99.69                          & B      & \ion{O}{vi}    & 2s\hses
      & \hsx  6p                &G & Kelly      & N, SMu\\
100.70, .90                    & A      & \ion{Mg}{vi}   &
      2p$^3$\,$^2$D$^{\rm o}$  & 3d\,$^2$F               &M & Kelly
      & N, SBu\\
101.49, .55                    & B      & \ion{Mg}{vi}   &
      2p$^3$\,$^2$D$^{\rm o}$  & 3d\,$^2$P               &M & Kelly
      & N, SBu,SMu\\
102.91, 103.08                 & A,B,P  & \ion{Ne}{viii} & 2p\hse\,$^2$P$^{\rm o}$  & 3s\,$^2$S\hse           &  & {\sc Nist} & SM,SMH,SW,D,R\\
103.09                         & (A,B,P)& \ion{Ne}{vii}  &
      2p\hsw\,$^1$P$^{\rm o}$  & 3p\,$^1$D\hse           &  & Kelly
      & N, SBu, blend with \ion{Ne}{viii}~103.08~\AA\\
104.81                         & B,P    & \ion{O}{vi}    & 2s\hses                  & \hsx  5p                &G & Kelly      & SMu,R\\
105.17                         & A,(B)  & \ion{Mg}{vii}  &
      2p$^3$\,$^1$D$^{\rm o}$  & 3s'\,$^1$D\hse          &  & Kelly
      & N, SMu, blend with \ion{Fe}{ix}~105.21~\AA\ noted by R\\
106.03, .08, .19               & P      & \ion{Ne}{vii}  &
      2p\hse\,$^3$P$^{\rm o}$  & 3d\,$^3$D\hse           &M & Kelly
      & N, SM,SW,D,R\\
(111.10, .16), .26             & A,B,P  & \ion{Ne}{vi}   &
      2p\hse\,$^2$P$^{\rm o}$  & 3p\,$^2$D\hse           &G & Kelly
      & N, SBu,SMu, blend with \ion{Ca}{x}~111.20\\
  & & & & & & & poss.~\ion{Mg}{vi} contribution noted by R\\
111.15                         & (A),B,P& \ion{Ne}{vii}  & 2p$^2$\,$^1$D\hses       & 3d\,$^1$P$^{\rm o}$     &M & Kelly      & N, blend with \ion{Ca}{x}~111.20, \ion{Ne}{vi}~111.16~\AA\\
  & & & & & & & poss.~\ion{Mg}{vi} contribution noted by R\\
111.55, .75, .86               & B,(A,P)& \ion{Mg}{vi}   & 2p$^3$\,$^4$S$^{\rm o}$  & 3s\hse\,$^4$P           &G & Kelly      & SB,R\\
(115.33), .39, (.52)           & A,B,P  & \ion{Ne}{vii}  & 2p\hse\,$^3$P$^{\rm o}$  & 3s\,$^3$S\hse           &M & Kelly      & R\\
115.82, .83                    & B      & \ion{O}{vi}    & 2s\hses                  & \hsx  4p                &G & Kelly      & SB,SMu,Ru\\
115.96                         & B      & \ion{Ne}{vii}  & 2p$^2$\,$^1$D\hses       & 3d\,$^1$D$^{\rm o}$     &M & Kelly      & N\\
(116.35), .42                  & B      & \ion{O}{vi}    & 2p                       & \hsx  5d                &  & Kelly      & N\\
116.69                         & B      & \ion{Ne}{vii}  & 2p\hsw\,$^1$P$^{\rm o}$  & 3d\,$^1$D\hse           &  & Kelly      & SMu,R\\
116.97, 117.22                 & A      & \ion{Mg}{vi}   &
      2p$^3$\,$^2$D$^{\rm o}$  & 3s'\,$^2$P              &M & Kelly
      & N, SMu, poss. \ion{Mg}{vi} contribution noted by R\\
(117.33), .40                  & B      & \ion{O}{vi}    & 2p                       & \hsx  5s                &  & Kelly      & N\\
(117.43), .66, (.78)           & P      & \ion{Mg}{vii}  & 2p$^3$\,$^3$S$^{\rm o}$  & 3s\hse\,$^3$P\hse       &  & Kelly      & N,Ru\\
(117.52), .64, (.81)           & P      & \ion{Mg}{vii}  & 2p$^3$\,$^3$P$^{\rm o}$  & 3p\hse\,$^3$P\hse       &  & Kelly      & N,Ru\\
120.20, .27, .33, .35, .42, .48& P      & \ion{Ne}{vii}  & 2p$^2$\,$^3$P\hse        & 3s\,$^3$P$^{\rm o}$     &  & Kelly      & N, blend with \ion{O}{vii}~120.33~\AA\ note by R\\
122.49, .69                    & B,P    & \ion{Ne}{vi}   &
      2p\hse\,$^2$P$^{\rm o}$  & 3d\,$^2$D\hse           &G & Kelly
      & N, SBu,SMu,SMH,D,R\\
123.59                         & P      & \ion{Mg}{vi}   &
      2p$^4$\,$^2$D\hse        & 3s$^{\rm iv}$\,$^2$D$^{\rm o}$&
      &Kelly & N, SMu,Ru\\
127.67                         & B,P    & \ion{Ne}{vii}  & 2p\hsw\,$^1$P$^{\rm o}$  & 3s\,$^1$S\hse           &  & Kelly      & SMu,R\\
129.78, .87                    & A,B,P  & \ion{O}{vi}    & 2p                       & \hsx  4d                &  & Kelly      & SMH,SB,R\\
130.31, .64                    & B      & \ion{Mg}{vi}   & 2p$^4$\,$^2$P\hse        & 3s$^{\rm v}$\,$^2$P$^{\rm o}$& &Kelly  & N\\
130.94, 131.09, .30            & A,B,P  & \ion{Mg}{vii}  &
      2p$^3$\,$^3$S$^{\rm o}$  & 3p\hse\,$^3$P\hse       &  & Kelly
      & N, SBu, blend with \ion{Fe}{viii}~130.94, 131.24~\AA \\
  & & & & & & & noted by R\\
132.22, .31                    & A,B    & \ion{O}{vi}    & 2p                       & \hsx  4s                &  & Kelly      & N\\
150.09, .12                    & B,P    & \ion{O}{vi}    & 2s\hses                  & \hsx  3p                &G & Kelly      & SMH,SB,D,R\\
           \noalign{\smallskip}
\hline
     \end{tabular}
\end{table*}

\section{Observations} 

\hh\ was observed with the {\it Chandra} LETG+HRC-S on September 27, 2000, with
an integration time of approximately 25\,ks. Flux was detected in the range
60~\AA--160~\AA. The spectrum is that of a hot photosphere, characterized by a
continuum with a large number of absorption lines from highly ionized species:
\ion{O}{v-vi}, \ion{Ne}{vi-viii}, and \ion{Mg}{v-viii}.  It rolls off at long
wavelengths due to ISM absorption. The maximum flux is detected near
110~\AA. Between 105~\AA\ and 100~\AA\ the flux drops because of photospheric
absorption from the \ion{O}{vi} edge caused by the first excited atomic
level. The edge is not sharp because of a converging line series and pressure
ionization. Below 100~\AA\ the flux decreases, representing the Wien tail of the
photospheric flux distribution. The complete spectrum with detailed line
identifications was presented in W04.

The \alab\ observation has been described in detail by Raassen \etal\ (2002) and
we describe it here only in brief.  \alp\ was observed with the LETG+HRC-S on
December 25, 1999 with an exposure time of 81.5\,ks,  including dead time
corrections to account for telemetry saturation during intervals of high
background.  The observation was designed such that the two stars were maximally
separated in the cross-dispersion axis, with the dispersion axis positioned
nearly perpendicular to the axis of the binary.  At the time of the observation,
the stars were separated by $16\arcsec$ on the sky. The spectra were extracted
with the standard CIAO bow-tie region, though the central two background regions
interfered with the stellar spectra and only the outer regions were used for
background subtraction.

The two Procyon observations studied here were obtained with the  LETG+HRC-S as
part of the {\it Chandra} on-orbit calibration programme and Emission Line
Project.  The observations were executed contiguously beginning on November 6,
1999 at 21:11:32 UT. The second observation began on 1999 November 16:59:48 UT.
The effective exposure times for these observations were 69,643s and 69,729s,
respectively, including dead time corrections.

Reduction of the HRC-S event lists for all the observations was initially based
on standard pipeline products.  Events were further filtered in pulse height in
order to remove background events. The final reduced first order spectra were
co-added in order to maximise the signal.  In the case of Procyon, we also
co-added the two separate observations.

\section{Photospheric model for \hh}

We use here a photospheric spectrum from a line blanketed non-LTE model
atmosphere constructed for \hh\ by W04. Model parameters are: \Teff=200\,000\,K,
\logg=8 [cm\,s$^{-2}$], and C=48\%, O=48\%, Ne=2\%, Mg=2\%, (mass fractions). Details
of model assumptions and calculations can be found in that reference and we
restrict ourselves here to those characteristics which are of immediate
relevance in our context.  This primarily concerns the NLTE model atoms for neon
and magnesium. They comprise 88 and 122 NLTE levels, connected with 312 and 310
radiative line transitions, respectively, in the ionization stages
\ion{}{iv-ix}. The final synthetic spectrum was computed considering fine
structure splitting of levels and multiplets assuming relative LTE populations
for levels within a particular term. We have tried to use the best available
data for level energies and line wavelengths, compiling them from several
sources. For the lines discussed here (Table~\,\ref{lines_tab}), we used the
following databases:

\noindent
(i) National Institute of Standards and Technology (NIST)\footnote{http://physics.nist.gov/},\\
(ii) {\sc Chianti} database (Young \etal 2003)\footnote{http://wwwsolar.nrl.navy.mil/chianti.html},\\
(iii) Kelly Atomic Line Database\footnote{http://cfa-www.harvard.edu/amdata/ampdata/kelly/kelly.html}.

\noindent
However, in order to assemble the complete model atoms, other sources were essential, too:

\noindent
(iv) Opacity Project (OP, Seaton \etal 1994) TOPbase\footnote{http://legacy.gsfc.nasa.gov/topbase/home.html},\\ 
(v) University of Kentucky Atomic Line List\footnote{http://www.pa.uky.edu/$^\sim$peter/atomic/}.

\section{Comparison of \hh\ with \ala, \alb, and \pp} 

We have performed a detailed comparison of the \hh\ photospheric absorption line
spectrum with the coronal emission line spectra of \ala, \alb, and \pp. We have
also used the model spectrum of \hh\ for this purpose. It turns out that not all
lines predicted by the model, particularly the weaker ones, are readily
identified in \hh, which is at least in part due to the S/N of the {\it Chandra}
spectrum. Another reason is heavy blending by lines from iron group elements,
which are not considered in the model used here. It was shown that
identification of weak lines suffers from iron and nickel line blends, which is
a problem because the accurate positions of the majority of lines from Fe-group
elements in the soft X-ray domain is unknown (W04).  The use of our synthetic
spectrum in addition to the \hh\ spectrum helps considerably to identify lines
in the coronal spectra.

Table~\,\ref{lines_tab} summarizes the results of our comparison. Lines from 65
multiplets of \ion{O}{vi}, \ion{Ne}{vi-viii}, and \ion{Mg}{vi-viii} are
identified in both, \hh\ (or its model) and in at least one of the considered
coronae. Many of these had already been identified in earlier solar work (Widing
\& Sandlin 1968; Freeman \& Jones 1970; Behring \etal 1972; Manson
1972; Malinovsky \& Heroux 1973; Acton et al.\ 1985) and by Raassen \etal (2002,
2003), but the majority represents new identifications.  Table~\ref{lines_tab}
also denotes lines or features seen in earlier solar spectra but which were
unidentified in the earlier work.  The identifications presented here can then
also be applied (either wholly or in part, allowing for blends) to these solar
spectra.  Many, but not all, of the tabulated lines have lower levels which are
either ionic ground states or metastable states (labeled G or M, respectively).
As an example how the spectra compare, we show in Fig.\,\ref{fig_84} the spectra
of Procyon and \hh\ in a wavelength region where a bunch of lines from two
\ion{Mg}{vii} and one \ion{Ne}{vii} multiplet is located. All three multiplets,
or at least some components of them, were identified by Raassen \etal (2002) in
Procyon. They are also clearly seen as absorption features in the \hh\
spectrum. Over this, we have plotted the model spectrum, degraded to the {\it
Chandra} spectral resolution, which can qualitatively reproduce the observed
line features. Placed at the top of this Figure we show the original,
non-degraded model spectrum, showing the diverse structure of the multiplets,
whose components are not entirely resolved in {\it Chandra} spectra, neither of
\hh\ nor of \pp.

Figure~\ref{fig_88} shows a detail from the spectra of \pp\ and \ala\ compared to
\hh\ in another wavelength interval. It displays some new line identifications
in the coronal spectra, see for example the 87.46~\AA\ resonance line of
\ion{Ne}{vii} in \ala. The strongest emissions in \ala\ stem from two
\ion{Ne}{viii} and \ion{Mg}{viii} doublets, identified already in Raassen \etal
(2003). But note that the \ion{Mg}{viii}~86.84~\AA\ component is blended with the
possibly stronger, newly identified \ion{Ne}{vii}~86.82~\AA\ line.

Some of the newly identified lines do blend with other lines used for coronal
diagnostics.  The emissivity of the \ion{Fe}{viii} lines at 130.94~\AA\ and
132.24~\AA\ in Procyon was computed by Raassen \etal (2002) using a
three-temperature model. They stress that these line strengths are strongly
underestimated, by factors 6 and 4 compared to the observation. The result of
their differential emission measure (DEM) model underestimates the emissivity
even more (factors 9 and 6). This can at least partially be explained by the
fact that two components of a \ion{Mg}{vii} triplet (at 130.94~\AA\ and
131.30~\AA) can contribute to the \ion{Fe}{viii} line emissivities. A similar
explanation may hold for the \ion{Fe}{ix}~105.20~\AA\ line, which also appeared
too weak in their model. It is blended with a \ion{Mg}{vii} singlet at 105.17~\AA.

Another example is the \ion{Mg}{viii}~74.86~\AA\ line observed in \ala\ and
\alb. Raassen \etal (2003) find that the line fluxes from their models are too
small by about 40\%. We think that the missing flux is contributed by a blend
with a new neon line located at almost the same wavelength,
\ion{Ne}{vii}~74.87~\AA. Detailed emission measure modeling, which is beyond the
scope of this paper, is needed to quantify these suggestions. Other blends with
previously identified emission lines in the coronae of Procyon and $\alpha$\,Cen
are indicated in Table~\ref{lines_tab}.

\section{Summary}

We have performed a detailed comparison of {\it Chandra} soft X-ray spectra from
the photosphere of the hottest known white dwarf, \hh, with the corona spectra
of \ala, \alb, and \pp. With the help of a detailed model spectrum for \hh\ we
have found that a large number of lines from multiplets of O, Ne, and Mg are
present in both the photospheric absorption line spectrum and the coronal
emission line spectra. In the coronal spectra we have newly identified lines
from about 40 multiplets of \ion{O}{vi}, \ion{Ne}{vi-vii}, and
\ion{Mg}{vi-viii}. Some of these lines are blends with previously known lines,
which are in use for diagnostic purposes, hence, their contribution to the line
flux must be considered in detailed spectral analyses.

\begin{acknowledgements}
Analysis of X-ray data in T\"ubingen is supported by the DLR under grant
50\,OR\,0201. JJD was supported by NASA contract NAS8-39073 to the {\it Chandra
X-ray Center}.
\end{acknowledgements}

\end{document}